\documentclass[sigconf]{acmart}

\AtBeginDocument{%
  }

\usepackage{threeparttable}
\usepackage{tabu}                     
\usepackage{multirow}                 
\usepackage{multicol}                 
\usepackage{float}                    
\usepackage{makecell}                 
\usepackage{booktabs}                 

\usepackage{graphicx}
\usepackage{float}
\usepackage{subfigure}
\usepackage[ruled,linesnumbered]{algorithm2e}
\usepackage{balance}

\begin{document}

\title{Next Interest Flow: A Generative Pre-training Paradigm for Recommender Systems by Modeling All-domain Movelines}

\author{Chen Gao}
\email{gaochen.gao@taobao.com}
\affiliation{%
  \institution{Alibaba Group}
  \city{Hangzhou}
  \country{China}}
  \orcid{0009-0005-6417-1428}

\author{Zixin Zhao}
\email{foriyte.zzx@taobao.com}
\affiliation{%
  \institution{Alibaba Group}
  \city{Hangzhou}
  \country{China}}
  \orcid{0009-0006-8638-7534}

\author{Lv Shao}
\email{shaolv.sl@taobao.com}
\affiliation{%
  \institution{Alibaba Group}
  \city{Hangzhou}
  \country{China}}
  \orcid{0009-0008-5389-6476}

\author{Tong Liu}
\email{yingmu@taobao.com}
\affiliation{%
  \institution{Alibaba Group}
  \city{Hangzhou}
  \country{China}}
  \orcid{0000-0003-2425-0357}

\renewcommand{\shortauthors}{Chen Gao et al.}

\begin{abstract}
  Click-Through Rate (CTR) prediction has long been dominated by discriminative paradigms that optimize local decision boundaries within candidate-specific subspaces. However, these models often fail to capture the global joint distribution and the continuous structural evolution of user intent across all-domain movelines. While generative approaches attempt to model global transition patterns, existing methods suffer from discretization-induced information collapse by remapping nuanced e-commerce signals into discrete linguistic or categorical spaces, failing to preserve the topological fidelity of interest trajectories. To overcome these limitations, we propose a novel generative pre-training paradigm that models user intent as a continuous evolutionary trajectory on a high-dimensional latent interest manifold, termed the Next Interest Flow (NIF). We introduce kinematic constraints to govern this flow: Interest Diversity is achieved via tangent space decomposition, while Evolution Velocity ensures trajectory smoothness through geodesic regularization. To bridge the objective mismatch between generative pre-training and discriminative fine-tuning, we propose a bidirectional alignment strategy to synchronize semantic spaces. Furthermore, we develop a Temporal Sequential Pairwise (TSP) mechanism to instill temporal causality within the discriminative framework. We present the All-domain Moveline Evolution Network (AMEN), a unified framework implementing this pipeline. Extensive experiments on a 6.7-billion instance industrial dataset and online A/B tests on Taobao validate AMEN's superiority, achieving +0.87pt AUC gain and +11.6\% CTCVR lift.
\end{abstract}





\maketitle

\section{Introduction}

Click-Through Rate (CTR) prediction is the fundamental optimization objective in industrial recommender systems. Despite the success of discriminative paradigms~\cite{DIN,SIM,DSIN,DIEN,CCN,DIAN,DIHN}, they are inherently constrained by learning hyperplane partitions within a restricted sample space defined by the preceding recommendation funnel. This approach primarily focuses on optimizing marginal decision boundaries, which limits the model’s capacity to reconstruct the underlying joint probability distribution of the all-domain user moveline or to capture the long-range structural dependencies of evolving preferences.

The emergence of generative paradigms offers a path toward modeling global transition patterns across the entire behavioral trajectory. However, current Large Language Model (LLM)-based~\cite{P5,M6Rec,HLLM} and ID-based methods~\cite{TIGER,HSTU,onerec,onerec_rep} often encounter a discretization-induced information collapse~\cite{VQVAE,HVQAE}. By remapping nuanced e-commerce signals into discrete linguistic tokens or categorical IDs~\cite{P5,TIGER}, these methods treat recommendation as the prediction of isolated, quantized endpoints. Such non-bijective transformations inevitably sacrifice the topological information inherent in the interest trajectory, thereby restricting the model’s ability to capture the fine-grained geometric context required for precise downstream estimation.

\begin{figure*}[htbp]
  \centering
  \includegraphics[width=0.97\linewidth]{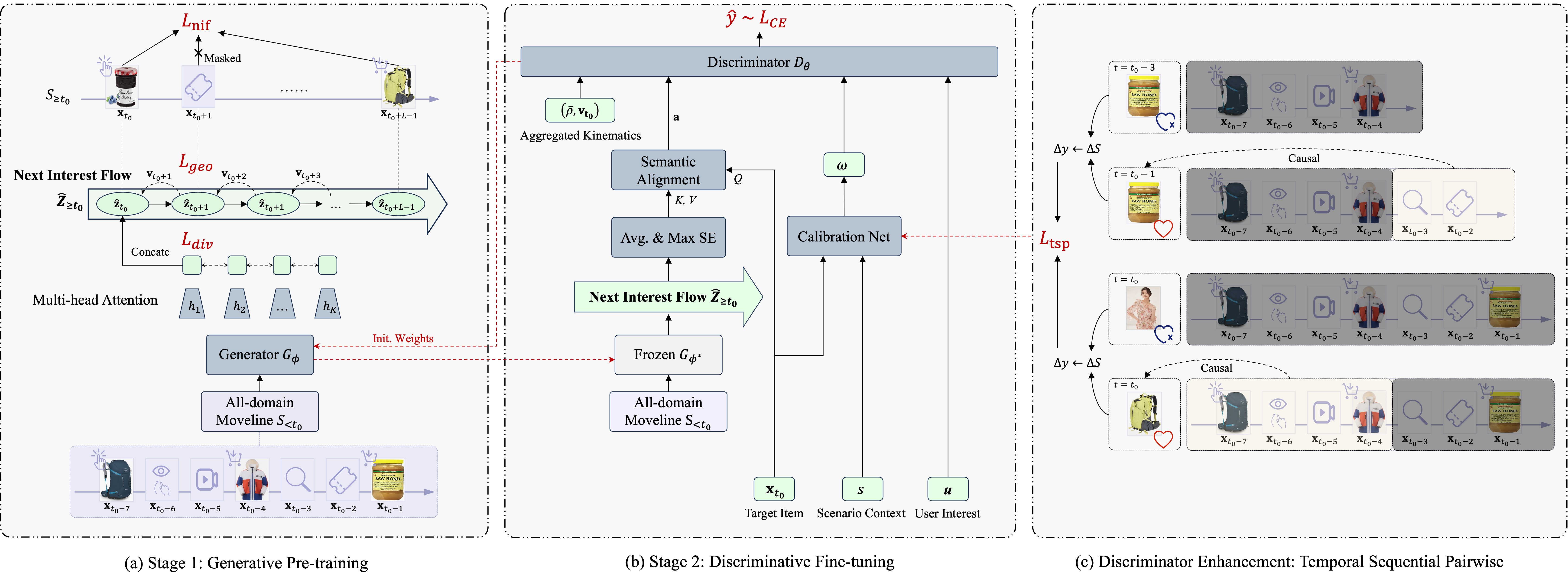} 
  \caption{The overall architecture of the All-domain Moveline Evolution Network (AMEN). (a) Stage 1: Generative Pre-training. (b) Stage 2: Discriminative Fine-tuning. (c) Discriminator Enhancement: Temporal Sequential Pairwise (TSP) Task.}
  \label{fig:framework}
\end{figure*}

In this work, we transcend these limitations by characterizing the user interest space as a high-dimensional Latent Interest Manifold $\mathcal{M}$. We propose a novel generative pre-training paradigm that models user intent as a continuous evolutionary trajectory $\gamma(t)$ on this manifold, termed the Next Interest Flow (NIF). By directly predicting the flow in the latent semantic space, NIF preserves the topological structure of user interactions and captures a universal Next Item connectivity. This provides a robust prior for long-tail and marketing-insensitive users, whose sparse discriminative signals are supplemented by the underlying manifold dynamics learned across the entire domain, consistent with the self-supervised representation paradigms in \cite{s3rec}.

To ensure the coherence and richness of the predicted trajectory, we govern the NIF through Kinematic Constraints derived from differential geometry:
\begin{itemize}
    \item \textbf{Interest Diversity} is achieved by decomposing the flow into multiple attention-based~\cite{attention} bases that span the local \textbf{tangent space} $\mathcal{T}_{\gamma(t)}\mathcal{M}$. This ensures the model captures a complete, non-redundant representation of the user’s multi-faceted intent.
    \item \textbf{Evolution Velocity} is defined by the first-order temporal derivative of the trajectory. By regularizing the evolution along a \textbf{geodesic path}, the model minimizes unstable accelerations, effectively filtering out stochastic noise and ensuring the temporal stability of the forecasted intent.
\end{itemize}

Integrating such a continuous generative flow into a discriminative predictor requires resolving the objective mismatch between distribution modeling and point-wise~\cite{pointwise1,pointwise2} classification. To this end, we propose a bidirectional alignment strategy: cross-stage weight initialization synchronizes the semantic origin of both stages, while a semantic alignment module dynamically projects the predicted trajectory onto the target item’s context during fine-tuning. Furthermore, to distinguish true causal state transitions from spurious correlations within the moveline, we enhance the discriminative foundation with a Temporal Sequential Pairwise~\cite{pairwise1,pairwise2} (TSP) mechanism to instill temporal causality.

We present the All-domain Moveline Evolution Network (AMEN), a unified framework implementing this paradigm. Our contributions are summarized as follows:
\begin{itemize}
    \item We propose a novel generative pre-training paradigm that models user intent as a continuous trajectory on a latent manifold (NIF), bypassing the scaling limits and information losses of discrete generative models.
    \item We develop geometric regularizers that govern interest diversity through tangent space decomposition and ensure evolution smoothness via geodesic constraints.
    \item We introduce a bidirectional alignment strategy to bridge the gap between generative and discriminative objectives, and a TSP mechanism to instill temporal causality.
    \item Extensive experiments on a 6.7-billion instance industrial dataset and online A/B tests on Taobao validate AMEN's superiority, achieving +0.87pt AUC gain and +11.6\% CTCVR~\cite{esmm} lift.
\end{itemize}

\section{Proposed Method}

AMEN is a unified framework comprising two stages: (i) a generative pre-training stage where a Transformer~\cite{attention} decoder $\mathcal{G}_\phi$ characterizes the Next Interest Flow (NIF) on a latent interest manifold, and (ii) a discriminative fine-tuning stage where a CTR predictor $\mathcal{D}_\theta$ integrates the pre-trained dynamics through bidirectional alignment and temporal causality constraints. The architecture is illustrated in Fig.~\ref{fig:framework}.

\noindent\textbf{Notation.} We characterize the user interest space as a smooth Riemannian submanifold $(\mathcal{M}, g) \subset \mathbb{R}^D$. A user's moveline is a sequence of discrete observations $\mathcal{S} = (\mathbf{x}_1, \dots, \mathbf{x}_T)$, where $\mathbf{x}_t \in \mathbb{R}^D$ is the embedding of the item at time $t$. The generator $\mathcal{G}_\phi$ has $K$ attention heads with head dimension $D_k = D/K$, where $K \leq D_k$. The prediction horizon is $L$.

\subsection{Stage 1: Generative Pre-training}

\subsubsection{Next Interest Flow}

The observed moveline traces a discrete path on $\mathcal{M}$. We define the \textbf{Next Interest Flow (NIF)} as the autoregressive extension of this path into the future latent space. For a given inference timestamp $t_0$, the NIF is represented by the trajectory matrix $\hat{\mathbf{Z}} = [\hat{\mathbf{z}}_{t_0}, \dots, \hat{\mathbf{z}}_{t_0+L-1}]^\top \in \mathbb{R}^{L \times D}$, where each row $\hat{\mathbf{z}}_t \in \mathbb{R}^D$ is a latent flow state at future timestep $t$.

To capture the transition dynamics, $\mathcal{G}_\phi$ generates the flow states autoregressively. Specifically, at each future timestep $t \in [t_0, t_0+L-1]$, the generator's input sequence $\mathcal{I}_t$ is constructed by concatenating the historical ground-truth observations with the previously generated flow states:
\begin{equation}
\label{eq:nif_gen}
\hat{\mathbf{z}}_t = \mathcal{G}_\phi(\mathcal{I}_t) = \mathcal{G}_\phi\bigl(\underbrace{\mathbf{x}_1, \dots, \mathbf{x}_{t_0-1}}_{\text{History (GT)}}, \underbrace{\hat{\mathbf{z}}_{t_0}, \dots, \hat{\mathbf{z}}_{t-1}}_{\text{Generated Flow}}\bigr), \quad \hat{\mathbf{z}}_t \in \mathbb{R}^D
\end{equation}
This hybrid input structure allows the model to extrapolate the interest trajectory based on its own predicted momentum while remaining anchored in the user's actual history. By operating in the continuous embedding space, NIF preserves the topological fidelity of interest trajectories and avoids the information collapse inherent in discrete tokenization.

The pre-training objective maximizes the mutual information between the predicted state $\hat{\mathbf{z}}_t$ and the actual future observation $\mathbf{x}_t$ via the InfoNCE~\cite{infoNCE} loss:
\begin{equation}
\label{eq:loss_nif}
\mathcal{L}_{\mathrm{nif}} = -\sum_{t=t_0}^{t_0+L-1} \log \frac{\exp\bigl(\langle\hat{\mathbf{z}}_{t}, \mathbf{x}_{t}\rangle / \tau\bigr)}{\sum_{\mathbf{x}' \in \{\mathbf{x}_t\} \cup \mathcal{N}_{t}} \exp\bigl(\langle\hat{\mathbf{z}}_{t}, \mathbf{x}'\rangle / \tau\bigr)}
\end{equation}
where $\mathcal{N}_t$ is the negative sample set and $\tau$ is the temperature.

\subsubsection{Kinematic Constraints}

We govern the evolution of NIF through two kinematic constraints derived from differential geometry to ensure trajectory consistency and representational richness.

\vspace{0.5em}
\noindent\textbf{Interest Diversity via Tangent Frame Orthogonality.}
Each flow state $\hat{\mathbf{z}}_t$ is the concatenation of $K$ head outputs $\{\mathbf{q}_k\}_{k=1}^K$. We interpret these as a local frame on the tangent space $\mathcal{T}_{\gamma(t)}\mathcal{M}$ and seek to maximize their spanning volume~\cite{kulesza2012determinantal}.

\begin{proposition}[Optimality of Orthogonal Frames]
\label{prop:diversity}
Let $\mathbf{W}_t \in \mathbb{R}^{K \times D_k}$ ($K \leq D_k$) be the frame matrix whose $k$-th row is the normalized output $\bar{\mathbf{q}}_k$. The representational volume $\mathrm{Vol}(\mathbf{W}_t) = \sqrt{\det(\mathbf{W}_t\mathbf{W}_t^\top)}$ is maximized if and only if $\, \mathbf{W}_t\mathbf{W}_t^\top = \mathbf{I}_K$.
\end{proposition}

\begin{proof}
Let $\mathbf{G} = \mathbf{W}_t\mathbf{W}_t^\top$ be the Gram matrix. Since $\|\bar{\mathbf{q}}_k\|_2 = 1$, the diagonal elements $G_{kk} = 1$. By Hadamard’s inequality, $\det(\mathbf{G}) \leq \prod_k G_{kk} = 1$, where equality holds if $\mathbf{G} = \mathbf{I}_K$.
\end{proof}

To optimize this, we employ a Frobenius relaxation~\cite{bansal2018can}:
\begin{equation}
\label{eq:loss_div}
\mathcal{L}_{\mathrm{div}} = \sum_{t=t_0}^{t_0+L-1} \left\| \mathbf{W}_t \mathbf{W}_t^\top - \mathbf{I}_K \right\|_F^2
\end{equation}
A per-state diversity score $\rho_t = 1 - \|\mathbf{W}_t\mathbf{W}_t^\top - \mathbf{I}_K\|_F^2 / K^2$ is derived for downstream use.

\vspace{0.5em}
\noindent\textbf{Evolution Smoothness via Geodesic Regularization.}
We impose a smoothness prior on the trajectory $\hat{\mathbf{Z}}$ to suppress stochastic noise and unstable accelerations~\cite{do1992riemannian,kass1988snakes,bronstein2017geometric}.

\begin{proposition}[Variational Optimality of Acceleration Penalization]
\label{prop:geodesic}
The natural cubic spline is the unique $C^2$ curve minimizing the bending energy $E[\gamma] = \int \|\ddot{\gamma}(t)\|_2^2 dt$. On a uniform temporal grid, this functional discretizes to the squared Frobenius norm of the second-order finite difference of the trajectory matrix $\hat{\mathbf{Z}}$.
\end{proposition}

\begin{proof}
By the Holladay theorem~\cite{ahlberg1967theory}, $E[\gamma]$ is minimized by a spline with $\ddot{\gamma}=0$ at boundaries. Discretizing via the central difference $\ddot{\gamma}(t_i) \approx \hat{\mathbf{z}}_{i+1} - 2\hat{\mathbf{z}}_i + \hat{\mathbf{z}}_{i-1}$ transforms the energy into $\|\boldsymbol{\Delta}_2 \hat{\mathbf{Z}}\|_F^2$, where $\boldsymbol{\Delta}_2$ is the second-order difference operator.
\end{proof}

The resulting geodesic regularization is:
\begin{equation}
\label{eq:loss_geo}
\mathcal{L}_{\mathrm{geo}} = \left\| \boldsymbol{\Delta}_2\, \hat{\mathbf{Z}} \right\|_F^2
\end{equation}

Additionally, we extract the first-order velocity $\mathbf{v}_t = \hat{\mathbf{z}}_t - \mathbf{x}_{t-1}$ (for $t=t_0$) or $\mathbf{v}_t = \hat{\mathbf{z}}_t - \hat{\mathbf{z}}_{t-1}$ (for $t > t_0$) to capture interest shifts.

\subsubsection{Pre-training Objective and Weight Initialization}

The Stage 1 joint objective is:
\begin{equation}
\label{eq:loss_stage1}
\mathcal{L}_{\mathrm{stage1}} = \mathcal{L}_{\mathrm{nif}} + \alpha \cdot \mathcal{L}_{\mathrm{div}} + \beta \cdot \mathcal{L}_{\mathrm{geo}}
\end{equation}
To ensure structural alignment between stages, $\mathcal{G}_\phi$ is initialized with weights from a pre-trained discriminative base model.

\subsection{Stage 2: Discriminative Fine-tuning}

The frozen generator $\mathcal{G}_{\phi^*}$ produces the trajectory matrix $\hat{\mathbf{Z}}$, the per-state diversity scores $\{\rho_t\}$, and velocity vectors $\{\mathbf{v}_t\}$.

\noindent\textbf{Semantic Alignment Module.}
To synchronize the generative flow with the candidate target item $\mathbf{x}_{t_0}$, we utilize target attention~\cite{DIN} to perform soft selection over the predicted interest evolution:
\begin{equation}
\label{eq:alignment}
\mathbf{a} = \mathrm{Attention}\bigl(\text{Query}=\mathbf{x}_{t_0},\; \text{Key}=\hat{\mathbf{Z}},\; \text{Value}=\hat{\mathbf{Z}}\bigr)
\end{equation}
This alignment projects the continuous manifold trajectory $\hat{\mathbf{Z}}$ onto the local decision context of the target item.

\noindent\textbf{Final Prediction.}
The aligned representation $\mathbf{a}$, historical user interest $\mathbf{u}$, aggregated kinematics ($\bar{\rho}$, $\mathbf{v}_{t_0}$), and the target item $\mathbf{x}_{t_0}$ are fed into a merge MLP to produce the main logit $\hat{y}_{\mathrm{main}}$. The final prediction is calibrated by the TSP score $\omega$:
\begin{equation}
\label{eq:prediction}
\hat{y} = \sigma\bigl(\hat{y}_{\mathrm{main}} + \omega\bigr)
\end{equation}

\subsection{Temporal Sequential Pairwise (TSP) Auxiliary Task}

To instill temporal causality, we select a diff sample from the same moveline at $t_1$ (within horizon $L$) having an opposite label. The calibration network in $\mathcal{D}_\theta$ computes scores $\omega_0$ and $\omega_1$. A BPR-style~\cite{bpr} objective is applied:

\begin{equation}
\label{eq:loss_tsp}
\mathcal{L}_{\mathrm{tsp}} = -\frac{1}{|\mathcal{P}|}
\sum_{(x_0, x_1) \in \mathcal{P}} \log \sigma\!\bigl(
\ell_{t_1} \cdot (\omega_1 - \omega_0)\bigr)
\end{equation}
where $\ell_{t_1}=2y_{t_1}-1 \in\{-1, +1\}$ is the label polarity 
indicator: $\ell_{t_1}=+1$ when the diff sample $x_1$ is positive, and $\ell_{t_1}=-1$ otherwise. The final fine-tuning loss is a weighted sum of a standard cross-entropy loss and the TSP loss:
\begin{equation}
\label{eq:loss_stage2}
\mathcal{L}_{\mathrm{stage2}} = \mathcal{L}_{\mathrm{CE}} + \lambda \cdot \mathcal{L}_{\mathrm{tsp}}
\end{equation}

\begin{figure*}[htbp]
  \centering
  \includegraphics[width=0.9\textwidth]{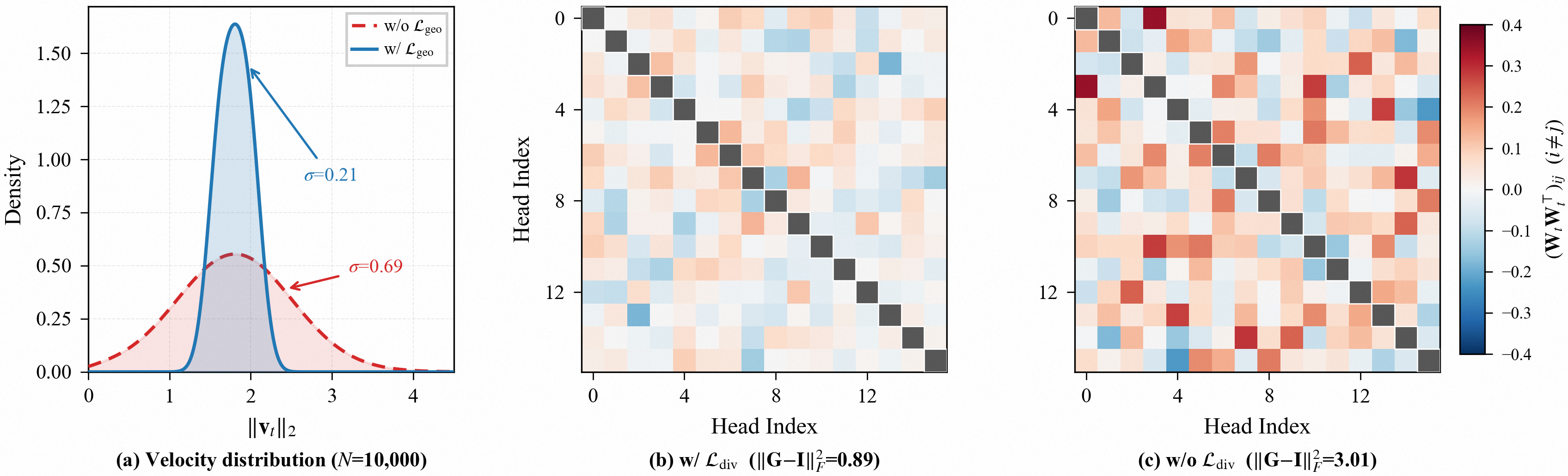} 
  \caption{Empirical effect of kinematic constraints. (a) Velocity norm distribution under $\mathcal{L}_{\mathrm{geo}}$. (b)--(c) Off-diagonal structure of the Gram matrix $\mathbf{W}_t\mathbf{W}_t^\top$ with and without $\mathcal{L}_{\mathrm{div}}$; diagonal entries are masked in gray. Near-zero off-diagonal values in (b) confirm the maintenance of a near-orthogonal tangent frame.}
  \label{fig:kinematic}
\end{figure*}

\section{Experiments}

\subsection{Setup}

\noindent\textbf{Dataset.} We evaluate AMEN on a Taobao industrial dataset with 6.7B training (180.7M users, 21.9M items) and 0.6B (23.2M users, 7.7M items) test instances (180.7M users, 21.9M items), using AUC as the primary metric.

\noindent\textbf{Baselines.} Baselines include discriminative models (Wide\&Deep~\cite{WideDeep}, DIN~\cite{DIN}, and MEDN, a strong internal discriminative baseline, which is an SIM~\cite{SIM} variant with extra contextual side info features and other interactions remaining unchanged) and generative paradigms (P5~\cite{P5}, TIGER~\cite{TIGER}).

We exclude general contrastive methods like CL4Rec~\cite{CL4Rec} and CoSeRec~\cite{CoSeRec} since TSP is conceptually orthogonal to them. The latter learn robust representations via augmentation, whereas TSP calibrates temporal misalignments and captures fine-grained sequential causality.

\noindent\textbf{Efficiency.} Generator and discriminator run independently, adding only +1\textasciitilde2\,ms online latency. While generative LLMs (e.g., TIGER/P5 0.5B) infer $\sim$60K samples daily per A100, AMEN processes 100M samples on a 12,800-core CPU cluster, demonstrating superior industrial scalability.

\noindent\textbf{Hyperparameters.} We conduct meticulous ablation studies over grouped hyperparameters, identifying the optimal configuration that achieves a trade-off between computational cost and performance: $\tau=0.1$, $\alpha=0.05$, $\beta=0.01$, $\lambda=0.1$, $L=48$ and $K=16$.

\subsection{Main Results and Ablations}

\begin{table}[htbp]
  \centering
  \caption{Offline Performance and Ablation Study. $\Delta$AUC is relative to the production baseline MEDN.}
  \label{tab:main_ablation_results}
  \small
  \resizebox{\linewidth}{!}{%
    \renewcommand{\arraystretch}{1.1}
    \setlength{\tabcolsep}{4pt}
    \begin{tabular}{llc}
    \toprule
    \textbf{Category} & \textbf{Model} & \textbf{AUC ($\Delta$AUC)} \\
    \hline
    \multirow{5}{*}{\textit{Baselines}} 
     & Wide\&Deep & 0.7216 ( -4.05pt ) \\
     & DIN & 0.7410 ( -2.11pt ) \\
     & P5 & 0.7565 ( -0.56pt ) \\
     & MEDN & 0.7621 ( - ) \\
     & TIGER & 0.7638 ( +0.17pt ) \\
     \hline
    \textit{Ours} & \textbf{AMEN (Full)} & \textbf{0.7708} ( \textbf{+0.87pt} ) \\
     \hline
    \multirow{6}{*}{\textit{Ablation}} 
     & \quad w/o NIF & 0.7694 ( +0.73pt ) \\
     & \quad w/o TSP & 0.7683 ( +0.62pt ) \\
     & \quad w/o Semantic Alignment & 0.7702 ( +0.81pt ) \\
     & \quad w/o Weight Initialization & 0.7698 ( +0.77pt ) \\
     & \quad w/o $\mathcal{L}_{\mathrm{div}}$ & 0.7697 ( +0.76pt ) \\
     & \quad w/o $\mathcal{L}_{\mathrm{geo}}$ & 0.7699 ( +0.78pt ) \\
    \bottomrule
    \end{tabular}%
  }
\end{table}

As shown in Table~\ref{tab:main_ablation_results}, AMEN achieves a significant +0.87pt AUC gain over the production baseline. Notably, AMEN substantially outperforms discrete generative models (P5, TIGER), validating that modeling intent as a continuous flow on $\mathcal{M}$ is superior to quantized symbolic prediction~\cite{VQVAE}. The ablation results highlight NIF and TSP as the primary performance drivers, while the kinematic regularizers ($\mathcal{L}_{\mathrm{div}}, \mathcal{L}_{\mathrm{geo}}$) provide crucial geometric refinement.

\subsection{Probing and Validation}

\noindent\textbf{Kinematic Validation.} We evaluate the trajectory properties over 10,000 users. Geodesic regularization ($\mathcal{L}_{\mathrm{geo}}$) causes the velocity norm $\|\mathbf{v}_t\|_2$ (Fig.~\ref{fig:kinematic}a) to concentrate sharply ($\sigma=0.21$), suppressing the erratic jumps ($\sigma=0.69$) seen in unconstrained models. This confirms that AMEN filters stochastic noise by constraining intent to a smooth manifold path. Furthermore, Fig.~\ref{fig:kinematic}b and Fig.~\ref{fig:kinematic}c show that under $\mathcal{L}_{\mathrm{div}}$, the tangent frame maintains near-orthogonality ($\|\mathbf{G}-\mathbf{I}\|_F^2=0.89$), maximizing representational volume as per Prop.~\ref{prop:diversity}. Removing $\mathcal{L}_{\mathrm{div}}$ leads to a $3.4\times$ increase in redundant correlation, resulting in subspace collapse.

\begin{figure}[htbp]
  \centering
  \includegraphics[width=0.88\linewidth]{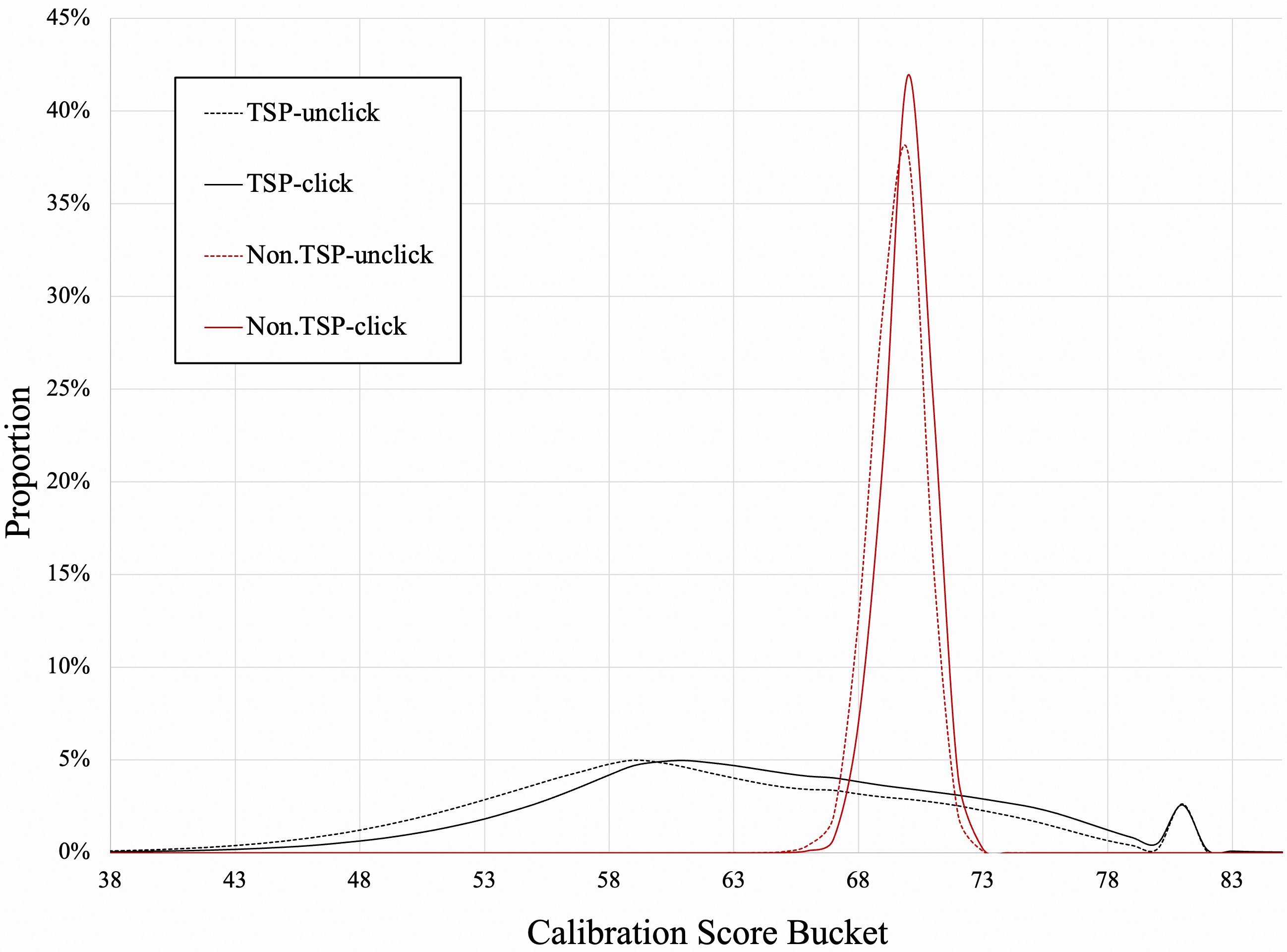}
  \caption{Probability density of TSP calibration scores.}
  \label{fig:gailvmidu}
\end{figure}

\noindent\textbf{Causal Calibration.} The TSP mechanism (Fig.~\ref{fig:gailvmidu}) successfully widens the margin of calibration scores $\omega$, instilling temporal-causal signals that traditional cross-entropy losses fail to capture.

\subsection{Online A/B Testing}

\begin{table}[htbp]
  \centering
  \caption{Online A/B testing results on Taobao.}
  \label{tab:abtest}
  \small
  \begin{tabular}{lccc}
    \toprule
    \textbf{Comparison} & \textbf{CTCVR} & \textbf{CTR} & \textbf{GMV} \\
    \midrule
    \multicolumn{4}{l}{\textit{Discriminative enhancements (TSP + alignment)}} \\
    Feeds: w/o NIF vs.\ MEDN & +11.6\% & Sta. & Sta. \\
    Floors: w/o NIF vs.\ MEDN & +4.2\% & +20.6\% & +20.1\% \\
    \midrule
    \multicolumn{4}{l}{\textit{+ Generative paradigm (full AMEN)}} \\
    Full vs.\ w/o NIF & +2.28\% & +0.98\% & +11.24\% \\
    \bottomrule
  \end{tabular}
  \begin{tablenotes}[flushleft]
    \footnotesize
    \item Sta.\ denotes statistically non-significant change.
  \end{tablenotes}
\end{table}

AMEN delivers +11.6\% CTCVR lift in the main Feeds scenario from discriminative enhancements, with additional +2.28\% CTCVR and +11.24\% GMV gains upon incorporating NIF, confirming real-world effectiveness.

\subsection{Statistical Significance} With the scale of 0.6 billion instances for offline tests, statistical significance testing trivially yield $p \ll 0.001$~\cite{significance_test,lin2013too}. We therefore emphasize practical significance to trasfer the AUC gain into substantial business impact. Online A/B testing is conducted during the biggest Taobao promotional event. The platform's rigorously stratified A/B framework confirms the robustness of our model. The reported lifts are sustained consistently across multiple days and diverse user segments.

\section{Conclusion}

This paper presents AMEN, a framework that addresses the limitations of discrete generative and local discriminative paradigms. By characterizing user intent as a continuous trajectory on a high-dimensional latent manifold, the Next Interest Flow preserves the topological fidelity of behavior sequences. Principled kinematic constraints, which are grounded in tangent frame orthogonality and geodesic smoothness, ensure the physical plausibility and diversity of the flow. Through bidirectional alignment and the TSP mechanism, we bridge the gap between distribution modeling and causal point-wise estimation. Extensive large-scale industrial experiments demonstrate that AMEN captures global evolution patterns, providing a robust and scalable paradigm for next-generation recommendation systems.

\bibliographystyle{ACM-Reference-Format}
\balance
\bibliography{article}
\end{document}